\documentclass[galaxies,article,submit,moreauthors,10pt,letterpaper]{mdpi_no_lineno}
\pdfoutput=1
\firstpage{1}
\articlenumber{x}
\doinum{10.3390/------}
\pubvolume{xx}
\pubyear{2016}
\copyrightyear{2016}
\externaleditor{Academic Editors:}
\history{Received: ; Accepted: ; Published:}

\theoremstyle{mdpi}
\newcounter{thm}
\newcounter{ex}
\newcounter{re}
\usepackage{soul}
\usepackage{microtype}
\newcommand\myurl[1]{\changeurlcolor{black}\url{#1}\changeurlcolor{blue}}
\makeatletter
\g@addto@macro{\UrlBreaks}{\UrlOrds}
\makeatother

\Title{Features of Structure and Absorption in the Jet-Launching Region of M87}

\Author{Wei Zhao $^{1,2,}$*, Xiaoyu Hong $^{1,2,3,4}$, Tao An $^{1,2}$, Xiaofeng Li $^{1,2,3,4}$,  Xiaopeng Cheng $^{1,2,3}$ and Fang~Wu~$^{1,2}$ }

\AuthorNames{Wei Zhao, Xiaoyu Hong, Tao An, Xiaofeng Li, Xiaopeng Cheng, and Fang Wu}

\address{%
$^{1}$ \quad Shanghai Astronomical Observatory, 80 Nandan Road, Shanghai 200030, China; xhong@shao.ac.cn (X.H.); antao@shao.ac.cn (T.A.); lixf@shao.ac.cn (X.L.); xcheng@shao.ac.cn (X.C.);  wufang@shao.ac.cn (F.W.)\\
$^{2}$ \quad Key Laboratory of Radio Astronomy, Chinese Academy of Sciences, Nanjing  210008,  China \\
$^{3}$ \quad University of Chinese Academy of Sciences, 19A Yuquanlu, Beijing 100049, China\\
$^{4}$ \quad ShanghaiTech University, 100 Haike Road, Pudong, Shanghai 201210, China}

\corres{Correspondence: weizhao@shao.ac.cn}

\abstract{M87 is one of the best available source for studying the AGN jet-launching region.
To enrich our knowledge of this region, with quasi-simultaneous observations using VLBA
at 22, 43 and 86~GHz, we capture the images of the radio jet in M87 on a scale within
several thousand $R_s$. Based on the images, we analyze the transverse jet structure and
obtain the most accurate spectral-index maps of the jet in M87 so far, then for the first time, 
we compare the results of the two analyses and find a spatial association between the jet 
collimations and the local enhancement of the density of external medium in the jet-launching region. 
We also find the external medium is not uniform, and greatly contributes to the free-free absorption 
in this region. In addition, we find for the jet in M87, its temporal morphology in the launching 
region may be largely affected by the local, short-lived kink instability growing in itself.}

\keyword{galaxies: individual (M87); galaxies: jets; galaxies: active;
radio continuum:galaxies}

\begin{document}

\section{Introduction}
\label{IN}
M87 is a nearby (D$\sim$16.7~Mpc, \citep{Har2009}) FRI radio galaxy 
which harbors a central black hole with {a mass reported} from 
2.4 to 7.2 $\times~10^9~\mathrm{M_{sun}}$ \citep{Har1994, Geb2011, Old2016}. 
Profiting from its proximity, the~angular resolution 
of ground-based Very Long Baseline Interferometry (VLBI) now is down to 
a few times of the angular size of schwarzschild radii ($R_s$) of the 
super massive black hole (SMBH) in M87. In~recent years, more and more 
facts of the vicinity of the SMBH in M87 are unveiled by high angular-resolution 
VLBI observations especially at millimeter wavelength. e.g.,~the accurate 
position of SMBH in M87 is constrained by~\cite{Had2011}; the central 
compact radio source in M87 is imaged directly by~\cite{EHT2019}.

M87 is a prototype to study the AGN jet-launching region as well as the vicinity 
of the SMBH. The~structure of the jet in its launching region has been determined 
with VLBI monitoring, that smears out the {variability details} by stacking multi-epoch 
observations at 22, 43 and 86~GHz \citep{Had2017, Wa2018, Kim2018}: the limb-brightened 
jet structure starts at a projected distance down to $7R_{s}$ to the SMBH, 
with an apparent opening angle wider than $100^{\circ}$ \citep{Had2016, Kim2018}; 
as the limb-brightened jet propagates outward, it has to experience multiple 
expansions followed with subsequent recollimations \citep{Wa2018, Kim2018} 
i.e., ``collimation regimes'' and finally reaches an equilibrium parabolic expansion 
in several thousand $R_s$, and~keeps the shape until $10^{5}R_{s}$~\citep{Asa2012, Had2013}. 
Besides the motion along the jet, the~jet flow rotates around the axis clockwisely, 
and the toroidal component of magnetic field is found close to the core by VLBI 
polarimetry \citep{Wa2018}. Additionally, the~counterjet is proved to be existent 
and with a limb-brightened structure as well \citep{Wa2018}.

In this work, we present images of the radio jet in M87 on a scale within 10~mas,
i.e., a few thousand $R_s$, captured by {very long baseline array (VLBA)} at 22, 
43 and 86~GHz {quasi-simultaneously}. Based on these images with high dynamical 
ranges, we analyze the transverse jet structure and obtain the most accurate spectral-index 
maps of the jet in M87 so far, and for the first time, we compare the results of the 
two analyses and find spatial associations between the structural features and the 
absorption features. We then make discussions on these associations.  We also discuss 
the cause of the temporal jet morphology at the observed epoch. Throughout this work, 
we use cosmological parameters 
$H_0 = 67.8~\mathrm{km~s^{-1}~Mpc^{-1}}$, $\Omega_{M} = 0.308$, $\Omega_{\Lambda}=0.692$, 
then 1~pc corresponds to 11.3~mas, and~$R_{s}\approx7\times10^{-4}~\mathrm{pc}=7.9~\mu \mathrm{as}$
if we take $\mathrm{M_{BH}}=7.2\times10^9~\mathrm{M_{sun}}$.

\section{Observations and Data~Reduction}
\label{OD}
In  28 March 2015, we observed M87 at 22~GHz ($\lambda=1.3~\mathrm{cm}$) 
and 43~GHz ($\lambda=7~\mathrm{mm}$) with VLBA. To~optimize the $u-v$ 
coverage, the~two observing frequencies were cycled in each of the 
observing runs. Six days later, we observed M87 at 86~GHz ($\lambda=3~\mathrm{mm}$) 
with VLBA. {For all three frequencies, some antennas were absent 
during the observations due to some technical problems. The~details of the
problems are described in the footnote of Table~\ref{Obslog}}. 
\begin{table}[H]
\centering 
\caption{Observations.}
\label{Obslog} 
\scalebox{0.95}[0.95]{
\begin{tabular}{ccccccc}
     \toprule  
\textbf{Date}  &  \textbf{Frequency} & \textbf{Telescopes }  & \textbf{On-Source Time}  & \textbf{Beam Size}\boldmath{$~^\mathrm{a}$}  & \boldmath{$I_\mathrm{peak}~^\mathrm{b}$ } & \boldmath{$\sigma~^\mathrm{c}$} \\
      & \textbf{ (GHz) }            &                & \textbf{(Min) }          & \textbf{(mas} \boldmath{$\times$} \textbf{mas, deg)}         & \textbf{(mJy~beam}\boldmath{$^{-1}$}\textbf{)           } &\textbf{(mJy~beam}\boldmath{$^{-1}$}\textbf{)}\\\midrule
2015 March 28 &  22.72       &  VLBA $^{\mathrm{d}}$                 & 124             &0.93$\times$0.45, 12.6                     &1054&0.360\\
           &  43.12       &  VLBA $^{\mathrm{d}}$                  & 124             &0.56$\times$0.23, 16.7                     &726&0.224\\
           2015 April 03 &  86.28       &  VLBA $^{\mathrm{e}}$       & 289             &0.24$\times$0.12, $-$17.7                 &501.4 & 0.217\\\bottomrule
\end{tabular}}\\ 

\begin{tabular}{@{}c@{}} 
\multicolumn{1}{p{\textwidth -.75in}}{\footnotesize Notes: \textsuperscript{a} The Synthesized beam FWHM size of the~images.  \textsuperscript{b} The peak flux density on the~images.  \textsuperscript{c} The root-mean-square noise of the~images.   \textsuperscript{d} Nine antennas were used, MK was absent due to the ``bad diskpack''~problem.  \textsuperscript{e} Seven antennas were used, HN and SC have no 86~GHz receiver while NL was ``disc~failure''.}
\end{tabular}
\end{table}

An automatic reference pointing control was used for the antenna-pointing.
5-min scans on the nearby bright source 3C 273 and 3C279 as fringe finders, 
delay and bandpass calibrators, were inserted during the observations, every 35
min for 22/43~GHz and 75 min for 86~GHz. The~received signals were 
sampled with 2-bit quantization and recorded with an aggregate data rate of 
$2048~\mathrm{Mbit~s}^{-1}$. See details of observations in Table~\ref{Obslog}.

The initial data calibration was made in AIPS (Astronomical Image 
Processing System) software package. For~22 and 43~GHz, we followed 
the standard procedure of VLBA data reduction. For~86~GHz, the~procedure
is slightly different: the global fringe-fitting was first performed 
on scans of 3C273 and 3C279 with point source models to derive time 
evolutions of the residual delay, rate, and~phase for each IF separately; 
the derived residual delay varies slowly with time, so the solutions 
of 3C 273 and 3C279's delay were used as a first-order approximation 
for M87's residual delay; then the global fringe-fitting was performed
twice on M87, the~first was performed with a point source model, 
the solution interval was set to 30~s and the threshold
of signal to noise ratio (SNR) was set to 3.0, to~avoid false
signals, a~tight delay and rate search window (10~nanoseconds and 20~mHz
for delay and rate respectively) were used, the~IFs are averaged 
to increase the SNR by a factor of 2.8; then we produce an initial
image, and~perform the second one with the CLEAN models obtained 
from the initial map to increase the fringe detection rate (the 
detection rate was increased slightly for 0.2$\%$). 
After calibration, multi-source data are split into single-source 
files and imported into DIFMAP for self-calibration and imaging. 
Self-calibration/Imaging loops are performed for multiple iterations 
to obtain images with high dynamical ranges. The~final image was 
produced in DIFMAP with natural~weighting.

\section{Results}
\label{RES}

\begin{figure}[H]
\centering
\includegraphics[angle=0, scale=0.52]{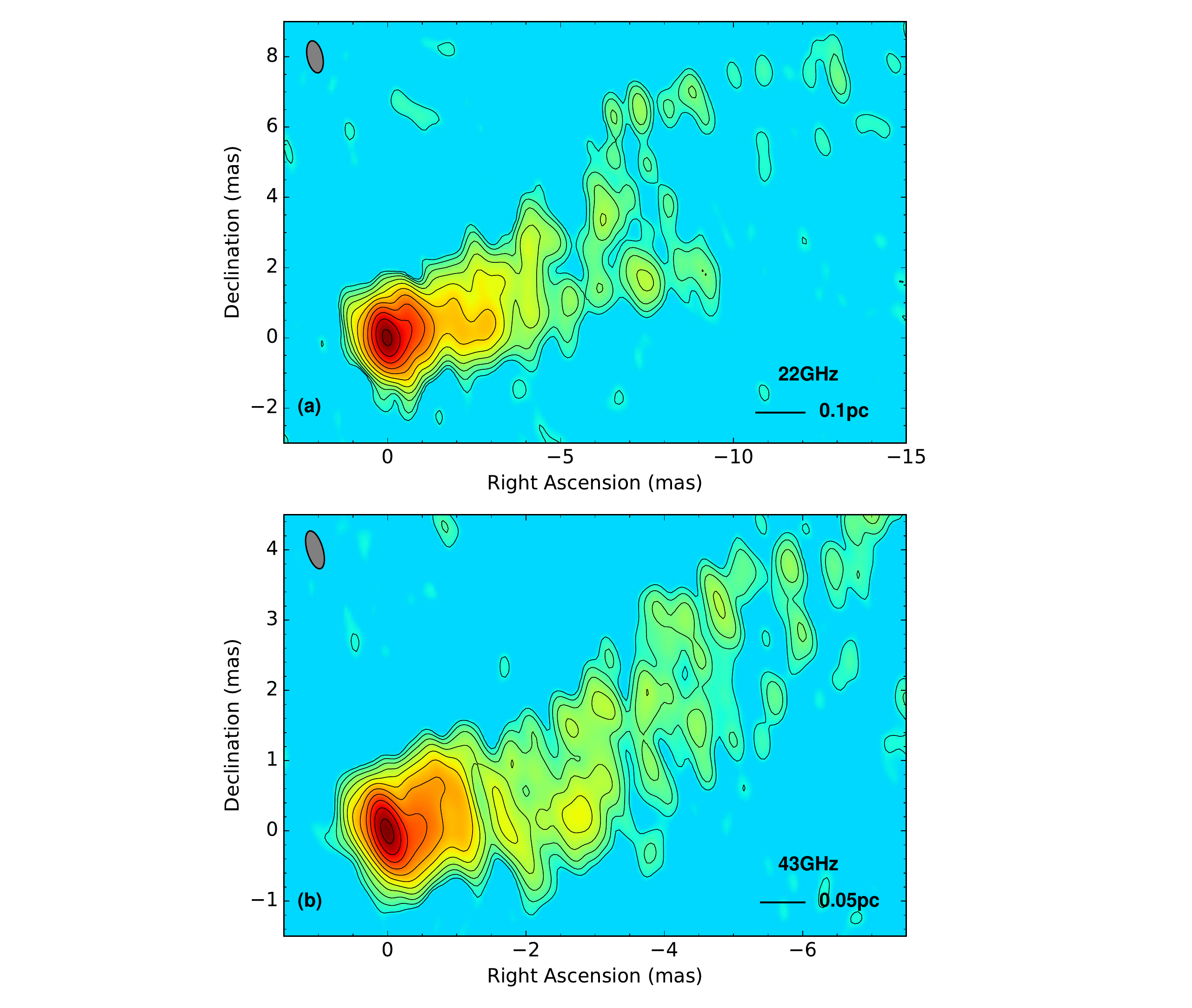}
\caption{22 and 43~GHz contour plots overlapping with pseudo-color images 
{of the same frequency respectively}, with~the synthesized beam shown at the top-left corner of each panel. 
Panel ({\textbf{a}}) shows the image of M87 at 22~GHz with a synthesized beam of 0.93~$\times$~0.45 
mas at a position angle of 12.6$^{\circ}$; contours in this image are from
($-$1, 1, 2, 4, 8...)$\times~\mathrm{0.86~mJy~beam}^{-1}$.
Panel ({\textbf{b}}) shows the image of M87 at 43~GHz with a synthesized beam of 0.56 $\times$ 0.23~mas
at a position angle of 16.7$^{\circ}$; contours in this image are from 
($-$1, 1, 2, 4, 8...)~$\times~\mathrm{0.54~mJy~beam}^{-1}$.}
\label{22_43}
\end{figure}

\begin{figure}[H]
\centering
\includegraphics[angle=0, scale=0.41]{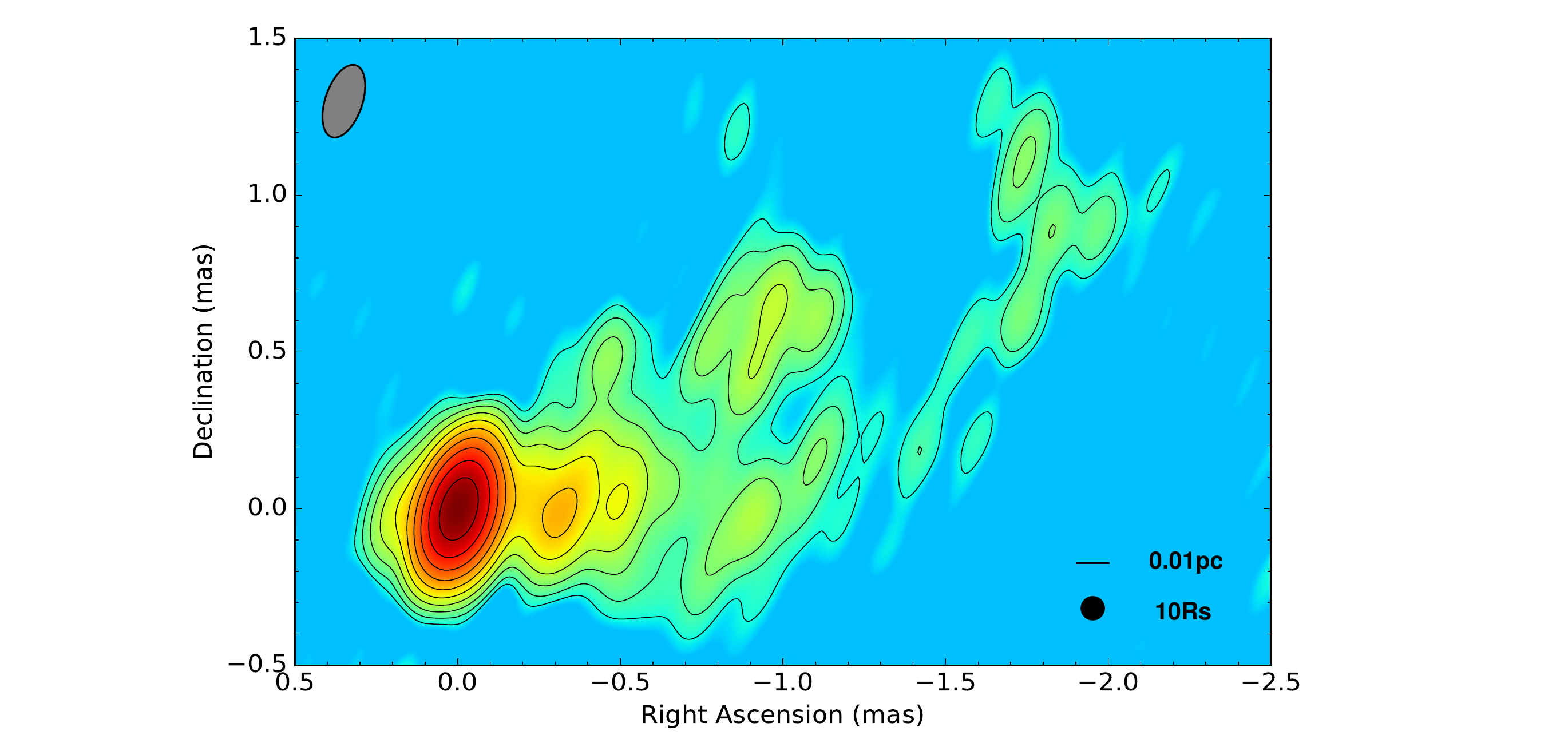}
\caption{86~GHz contour plot overlapping with pseudo-color image {of the same frequency} of M87; 
contours are from ($-$1, 1, 2, 4, 8...)~$\times~\mathrm{0.60~mJy~beam}^{-1}$.
The synthesized beam of 0.24 $\times$ 0.12~mas at a position angle of 
$\mathrm{-17.7}^\circ$ is shown at the top-left corner.}
\label{86_86}
\end{figure}

We show the resulting VLBI images in Figures~\ref{22_43} and ~\ref{86_86} as contour plots overlapping with pseudo-color 
images {of the same frequency respectively}. We list the parameters of the images (e.g., synthesized 
beam size, peak flux density and root-mean-square noise $\sigma$) 
in Table~\ref{Obslog}. Some structural features in the jet are 
apparently seen, e.g.,~on the 22~GHz image, the~jet shrinks locally 
between 1--2~mas from the center of the image, and~the northern 
limb is interrupted between 5 and 7~mas from the center of the image; 
on the 43~GHz image, the~jet bends southward between 1--2~mas from 
the center of the image; on the 86~GHz image, the~northern limb 
looks rather ragged, while the shape of its southern counterpart 
is far better defined; emission at the northeast of the most 
luminous region which is probably the hint of the counter jet, 
is seen on images at all three frequencies~.

\unskip

For the further analysis, we fit the most luminous region on the 
images as an elliptical Gaussian with procedure MODELFIT in DIFMAP, 
and refer it as ``core'' hereafter. We list the fitting results of the 
parameters in Table~\ref{Core} with their uncertainty estimated as 
in~\cite{Lee2008}. We also estimate the brightness temperature of 
the core as in~\cite{GUI1996}. The~brightness temperature of the 
core is estimated to be at the magnitude of $10^{10}~\mathrm{K}$ for 
all three frequencies. This is the same as given in~\cite{Kim2018, Had2016},
which is nearly one order of magnitude lower than 
the equipartition brightness~temperature.
 
\begin{table}[H]
\centering
\caption{Core~Component.}
\label{Core}
\scalebox{0.86}[0.86]{

\begin{tabular}{cccccccc}\toprule
\textbf{Frequency}& \textbf{Flux Density} \boldmath{$^a$}& \textbf{Radius} \boldmath{$^b$}& \textbf{Theta} \boldmath{$^c$}       & \textbf{Major Axis} \boldmath{$^d$}& \textbf{Axial Ratio} \boldmath{$^e$}& \textbf{Phi} \boldmath{$^f$}& \textbf{Brightness Temperature} \boldmath{$^g$}   \\
  \textbf{(GHz)}  & \textbf{(mJy)}           & \boldmath{$(\mathrm{\mu as})$}      & \textbf{(deg) }        & \boldmath{$(\mathrm{\mu as}) $}       &                &\textbf{ (deg)}  & \boldmath{$(10\mathrm{^{10}K})$ } \\\midrule
   22    &1605 $\pm$ 195     &19.9 $\pm$ 2.1 & $-$80.8 $\pm$ 6.1 & 479.2 $\pm$ 1.1 & 0.84           & 43.4   & 3.1  \\ 
   43    &1016 $\pm$ 122     & 3.6 $\pm$ 0.4 & $-$142.8 $\pm$ 5.8 & 261.2 $\pm$ 0.2 & 0.62           & 20.2   & 2.3  \\
   86    &580 $\pm$ 58       & 3.6 $\pm$ 0.3 & $-$44.7 $\pm$ 4.4 &  73.0 $\pm$ 0.2 & 0.52           & 31.8   & 5.0  \\ \bottomrule
\end{tabular}}\\

\begin{tabular}{@{}c@{}} 
\multicolumn{1}{p{\textwidth -.75in}}{\footnotesize Notes: $^a$ Integral flux density of~Gaussians. $^b$ Distance of Gaussians to the center of the~images. $^c$ Position angle of Gaussians referring to the~north. $^d$ Size of Major axis of~Gaussians. $^e$ Axial ratio of~Gaussians. $^f$ Position angle of major axis of Gaussians referring to the~north. $^g$ Brightness temperature of~Gaussians.}
\end{tabular}

\end{table}
\unskip

\subsection{Transverse Jet~Structure}
\label{SSJM}

\begin{figure}[H]
\begin{center}
\includegraphics[angle=0, scale=0.720]{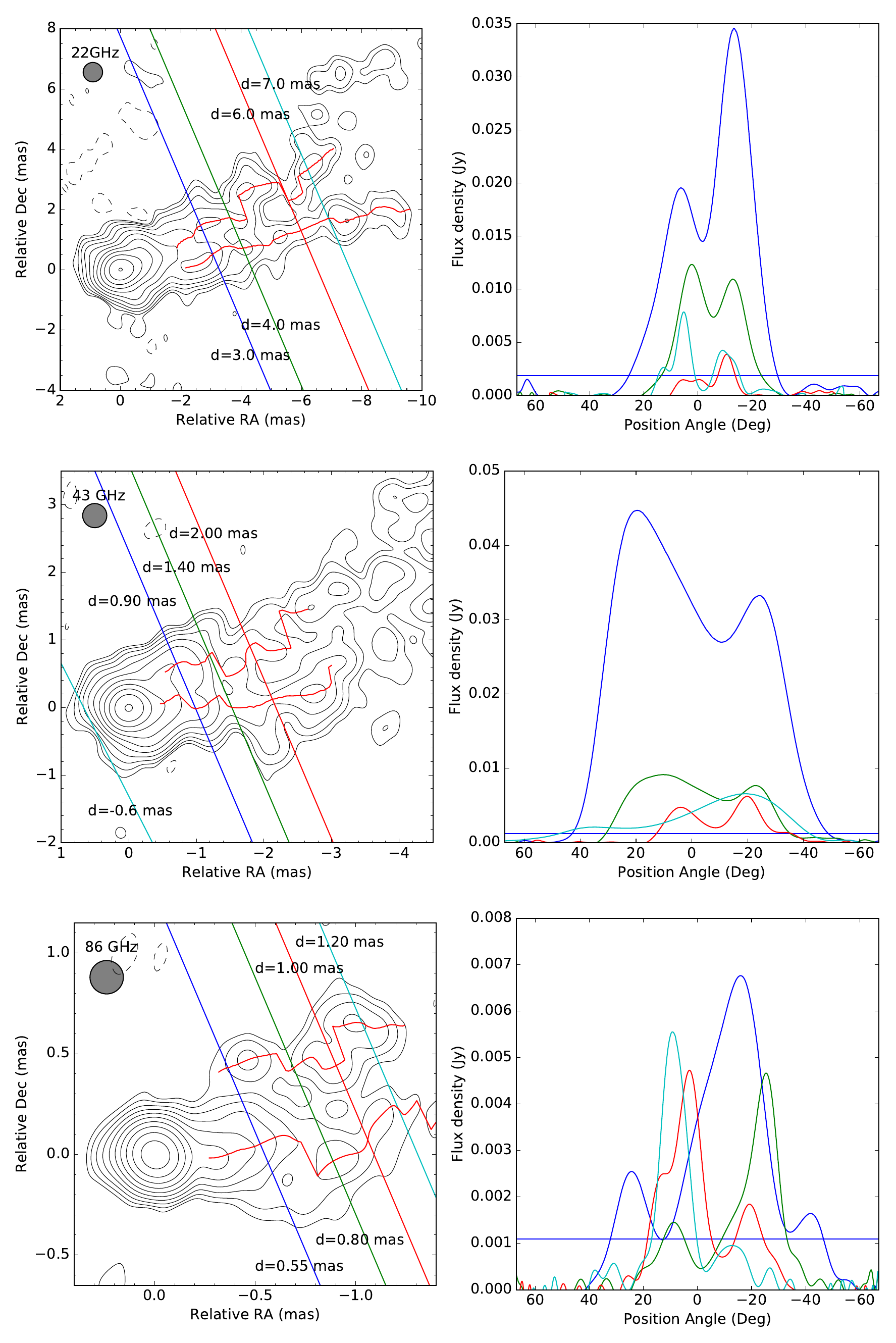}
\caption{The left panels shows the images restored with circular beams
as contours. The~restored beams are  
shown at the top-left corner of each panel. 
Locations of sample slices are presented as lines 
superimposed on the contours. The~ridgeline is presented 
as red curves superimposed on the contour plots.
The right panels present sample slices, showing 
the intensity as functions of position angles PAs referring 
to the jet axis. The~northern limb is at positive values, 
while its southern counterpart is at negative values. The~blue
horizontal lines represent 5$\sigma$ of the restored images.}
\label{flux2d}
\end{center}
\end{figure}

In this part, we use intensity slices to extract details of the jet and 
to analyze the transverse jet structure evolving with core distance $d$. 
To improve the angular resolution in the direction transverse to the jet 
with the least distortion of the images, in~advance of slicing, we restored 
the images with circular Gaussian beams whose diameter equal to the geometric 
mean of the length of major and minor axis of their synthesized beams, 
which is 0.65, 0.36 and 0.17~mas for the image at 22, 43 and 86~GHz respectively.
The restored images are shown as contours in the left panels of Figure~\ref{flux2d}. 
Then we slice the restored images in a direction perpendicular to the 
approximate overall jet axis, which is $-67^{\circ}$ respect to the 
north (referred as ``jet axis'' hereafter). For~the 22~GHz image, we made 
slices between $d=$~~2.00 and 10.00~mas, at~an interval of 0.05~mas; for 
the 43~GHz image, we made slices between $d$ = 0.10 and 3.00~mas, at~an 
interval of 0.05~mas; for the 86~GHz image, we made slices between $d$ =~0.10 
and 1.40~mas, at~an interval of 0.01~mas. The~right panels of Figure~\ref{flux2d} 
present the sample slices, showing the intensity as a function of 
position angles referring to the jet axis ($\mathrm{PAs}$, the~northern 
limb is at positive values, while its southern counterpart is at negative 
values). Locations of these sample slices are presented as lines 
superimposed on the contours in the left panels of Figure~\ref{flux2d}.

In our intensity slices, at~43 and 86~GHz, the~northern limb represents 
as a well defined hump since $d$~=~0.65 and 0.50~mas respectively, while its 
southern counterpart since $d$ = 0.45 and 0.24~mas respectively. The~counter 
jet emission also seems bifurcated on the restored 43~GHz image, and~a slice 
at $d$ = 0.60~mas shows this bifurcated structure clearly (see the cyan line in 
the second row of Figure~\ref{flux2d}). We also recover the ridgeline of 
the limb-brightened jet in M87 with the intensity slices. The~ridgeline is 
presented in the left panels of Figure~\ref{flux2d} as red curves 
superimposed on the contour plots. As~our result shows, both limbs are 
tortuous within $d=$~1.30~mas; as going outward, the~northern limb is still 
tortuous, while its southern counterpart becomes~smooth. 

\begin{figure}[H]
\begin{center}
\includegraphics[angle=0, scale=0.5]{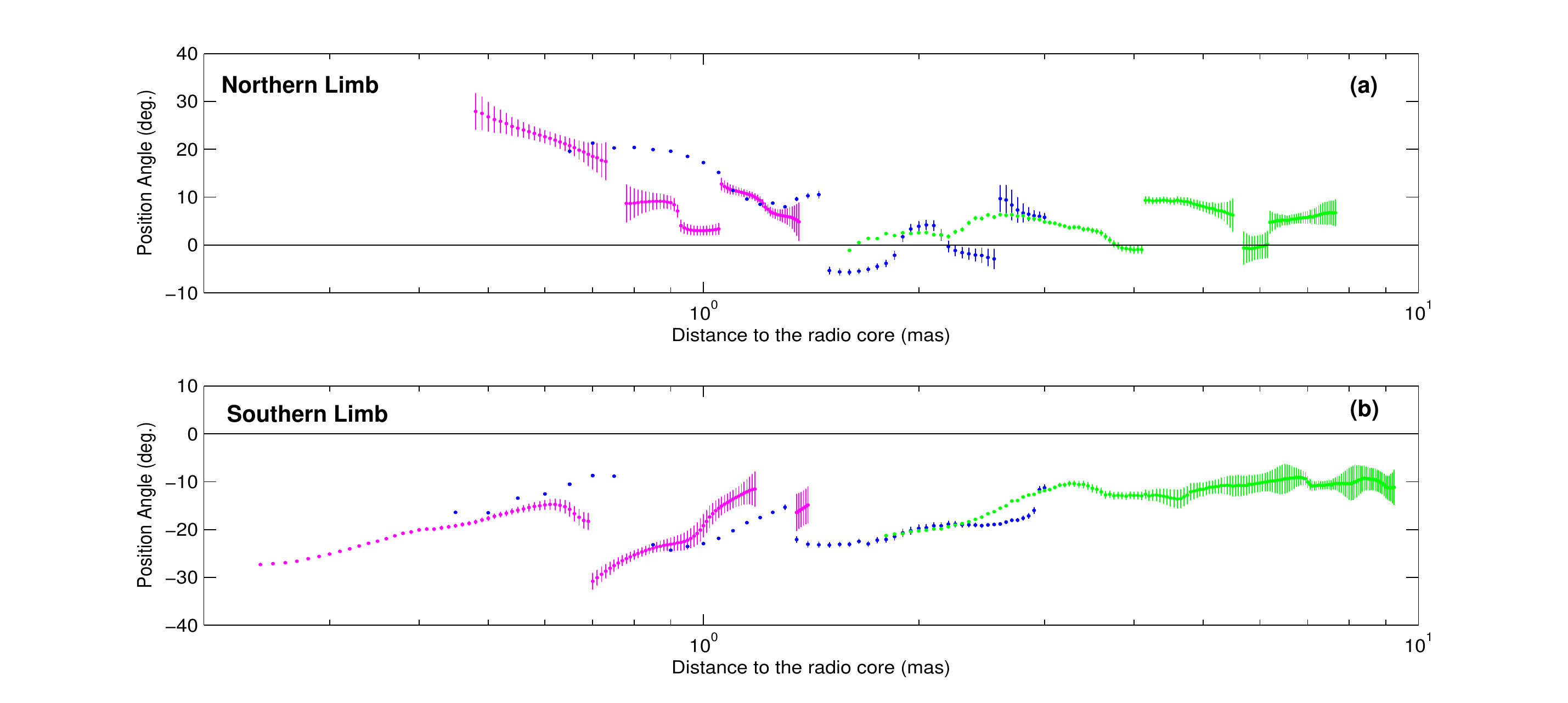}
\caption{(\textbf{a}) Position angles of the northern limb in M87 as functions of core distance $d$.
(\textbf{b}) Position angles of the southern limb in M87 as functions of core distance $d$.
The colors represent the measured frequencies, magenta for 86~GHz, 
blue for 43~GHz and green for 22~GHz. The~uncertainty of $\mathrm{PA}$ is estimated as 
$\frac{20^{\circ}}{\mathrm{SNR}}$, in~which $20^{\circ}$ is a typical width of the jet 
limb, and~SNR is the ratio between the intensity peak and the $\sigma$ of images. 
only peaks with intensity above 5$\sigma$ are presented to
make sure the result is reliable. }
\label{n-s}
\end{center}
\end{figure}

We present $\mathrm{PAs}$ of the intensity peak of both limbs as a function of $d$ 
in Figure~\ref{n-s}. The~general trend is both limbs approach to the jet 
axis gradually as going outward, but~there are some local structures seen on both 
limbs. We also notice that, if~the structure measured at higher frequency is 
moved outward for 0.1 or 0.2~mas, then it looks quite consistent with that measured 
at the adjacent lower frequency. This apparent shift could possibly be attributed to a
resolution effects which occur if there are gradients in the intensity for a jet. 
This is a combination of gradients in intensity, coupled with angular resolution 
depending on frequency.
Within $d=$~1.30~mas, at~86~GHz, the~northern limb is deflected northward at $d$ = 1.05~mas, 
while at 43~GHz, the~corresponding deflection is seen at $d=$~1.30~mas; at 86~GHz, the~southern limb 
veers southward at $d$ = 0.62 and 1.18~mas respectively, while at 43~GHz the corresponding 
changes are seen at $d$ = 0.75 and 1.30~mas respectively. Beyond~$d=$~1.30~mas, the~northern 
limb becomes fragmentized, and~its orientation changes abruptly at least at three locations 
($d=$~1.45, 2.55 and 4.55~mas), a~wiggle is seen between $d$ = 1.70 and 2.30~mas, and~an 
interruption is seen around $d$ = 6.00~mas; on the contrary, the~southern limb evolves 
smoothly as going outward, only a mild wiggle is seen between $d$ = 3.00 and 4.00~mas,
and its $\mathrm{PA}$ tends to be stable at $\sim-10^{\circ}$ since $d=$~4.00~mas.

\unskip

\begin{figure}[H]
\begin{center}
\includegraphics[angle=0, scale=0.5]{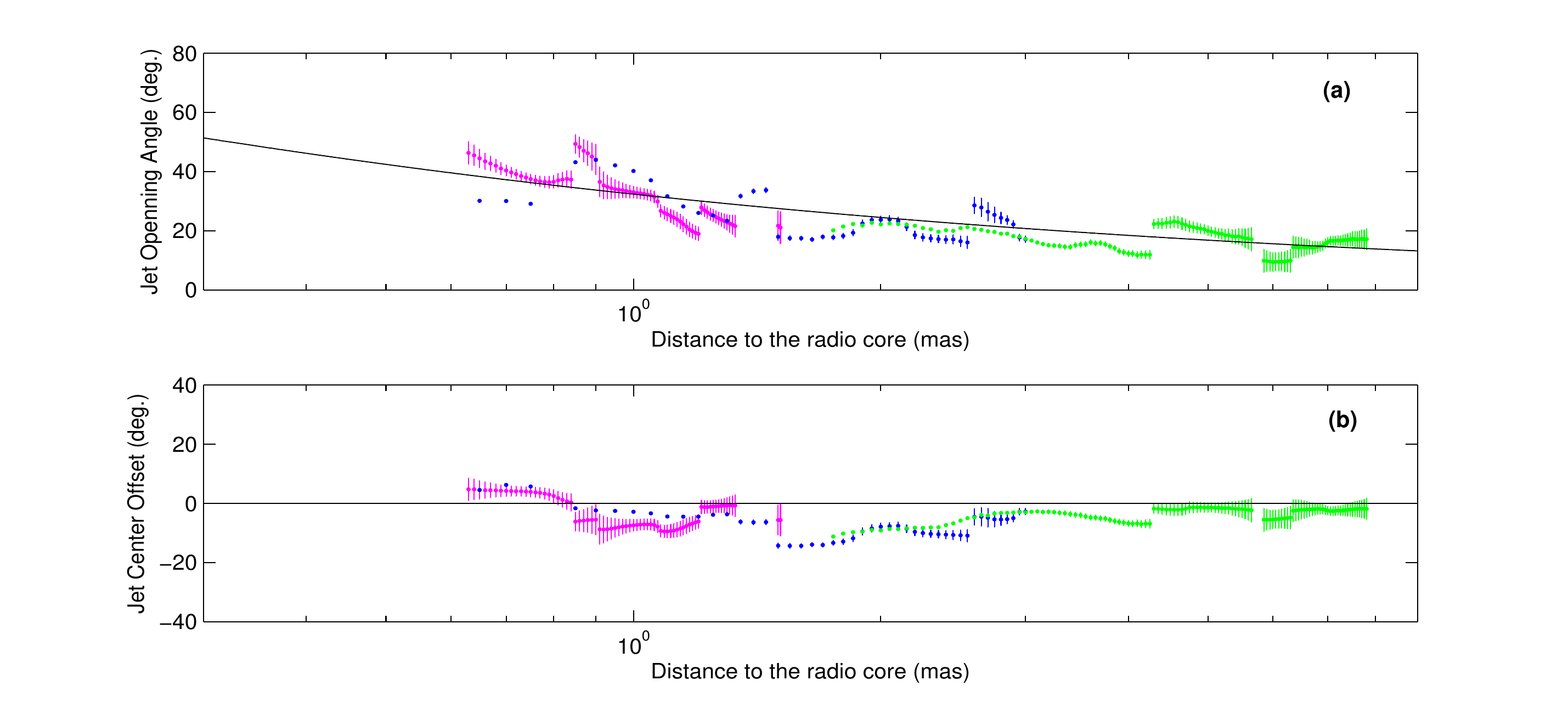}
\caption{(\textbf{a}) The apparent opening angle $\psi_{app}$ as functions of core distance $d$. 
Parabolic collimation profile determined by Asada \& Nakamura is shown for reference.
(\textbf{b}) The jet center offset $\delta_{app}$ as functions of core distance $d$. 
The jet axis is are shown for reference.
The colors represent the measured frequencies: magenta for 86~GHz, blue for 43~GHz 
and green for 22~GHz. The~uncertainties of $\psi_{app}$ and $\delta_{app}$ are 
estimated as $\sqrt{\Delta\mathrm{PA^2_{North}}+\Delta\mathrm{PA^2_{South}}}$.
Only peaks with intensity above 5$\sigma$ are presented to make sure the 
result is~reliable.}
\label{trans}
\end{center}
\end{figure}

In the following analysis, we move the jet structure measured at 86~GHz outward for 0.15~mas, 
and the structure measured at 22~GHz inward for 0.25~mas to eliminate the effect brought by 
different angular resolutions. We define the apparent jet opening angle $\psi_{app}$ as the 
difference of $\mathrm{PAs}$ of two limbs. The~panel (a) of Figure~\ref{trans} shows 
$\psi_{app}$ as a function of $d$. The~parabolic collimation profile, $\psi_{app}\propto r^{~0.58}$, 
determined by~\cite{Asa2012,Had2013}, is also presented for reference. 
The overall jet generally evolves around the parabolic profile, but~some local structural 
features are clearly seen. So we divide the jet into six segments, and~each segment includes 
an expansion and a following recollimation: segment 1 includes an quick expansion between 
$d$ = 0.75 and 0.90~mas, followed by a gradual recollimation between $d=$~0.90 and 1.30~mas; 
segment 2 includes an expansion between $d=$~1.30 and 1.45~mas, followed by a quick recollimation 
between $d=$~1.45 and 1.50~mas; segment 3 includes an gradual expansion between $d=$~1.50 and 2.00~mas, 
followed by a gradual recollimation between $d=$~2.00 and 2.55~mas; segment 4 includes a quick 
expansion between $d$~=~2.55 and 2.60~mas, followed by a slow recollimation between $d=$~2.60 and 3.35~mas; 
segment 5 includes an expansion between $d$ = 3.35 and 3.65~mas, followed by a gradual recollimation 
between $d=$~3.65 and 4.25~mas; segment 6 includes an expansion between $d=$~4.25 and 4.60~mas, 
followed by a recollimation beyond $d=$~4.60~mas. We define the apparent offset of jet center 
$\delta_{app}$ as the arithmetic average of the $\mathrm{PAs}$ of two jet limbs as well. 
The panel (b) of Figure~\ref{trans} shows $\delta_{app}$ as a function of $d$, and~the jet axis 
is presented as a black solid line for reference. As~the Figure shows, the~jet bends southward 
since $d=$~0.75~mas, the~offset reaches its maximum at $d=$~1.45~mas (where it is most clearly seen 
on the 43~GHz image), where the northern limb veers south, and~gradually returns to zero around $d=$~3.00~mas.

By comparing the behaviors of each limbs with $\psi_{app}$ evolving with $d$, 
we find that, structural features including the expansion and recollimation in segment 2, 3, 4 and 6, 
result from behaviors of the northern limb only, while the features in segment 1 and 5 result from 
behaviors of both limbs. By~further comparing with earlier works, we notice that the gradual recollimations 
in segment 1 between $d=$~0.90 and 1.30~mas and in segment 5 between $d=$~3.65 and 4.25~mas, 
spatially associate with recollimations in the first and second collimation regime (at $d=$~1.1 and 3.7~mas respectively) suggested by~\cite{Wa2018}. So they could possibly be the representations 
of these persistent recollimation features at our observed~epoch.

\subsection{Spectral-Index~Distribution}
\label{SSSP}
In this part, we present the spectral-index distribution in the jet
calculating between 22 and 43~GHz. To~align the VLBI images at different 
frequencies, we employ an algorithm which is first introduced by~\cite{CG2008}. 
It is based on the premise that the positions of optically thin regions 
to synchrotron radiation are not affected by absorption effects, then 
images can be aligned by finding out the maximum of 2D cross-correlation
of the optically thin regions. This method is now commonly used \citep{Fre2015,Mol2016,Pla2019} 
and was once employed by~\cite{Par2019} to obtained the RM map of jet in~M87.

\begin{figure}[H]
\begin{center}
\includegraphics[angle=0, scale=0.65]{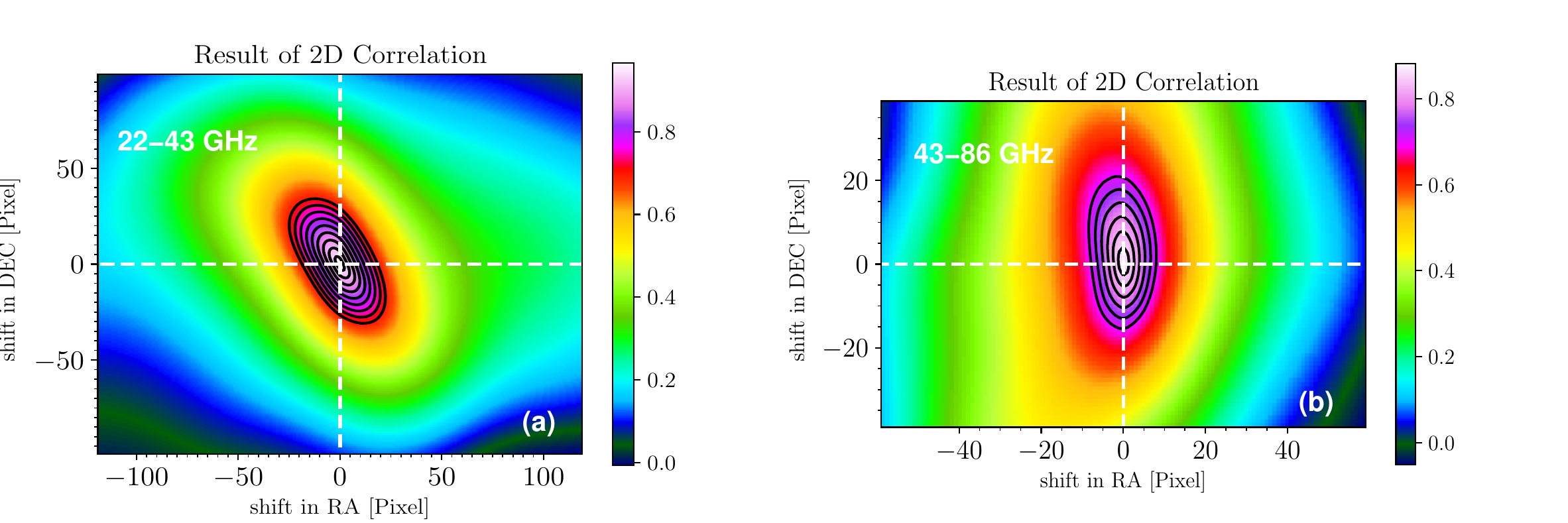}
\caption{Panel (\textbf{a}) shows 2D cross-correlation of the optically thin regions of 22 and 43~GHz images.
Panel (\textbf{b}) 2D cross-correlation of the optically thin regions of 43 and 86~GHz images.}
\label{2dc}
\end{center}
\end{figure}

\begin{figure}[H]
\centering
\includegraphics[angle=0, scale=0.8]{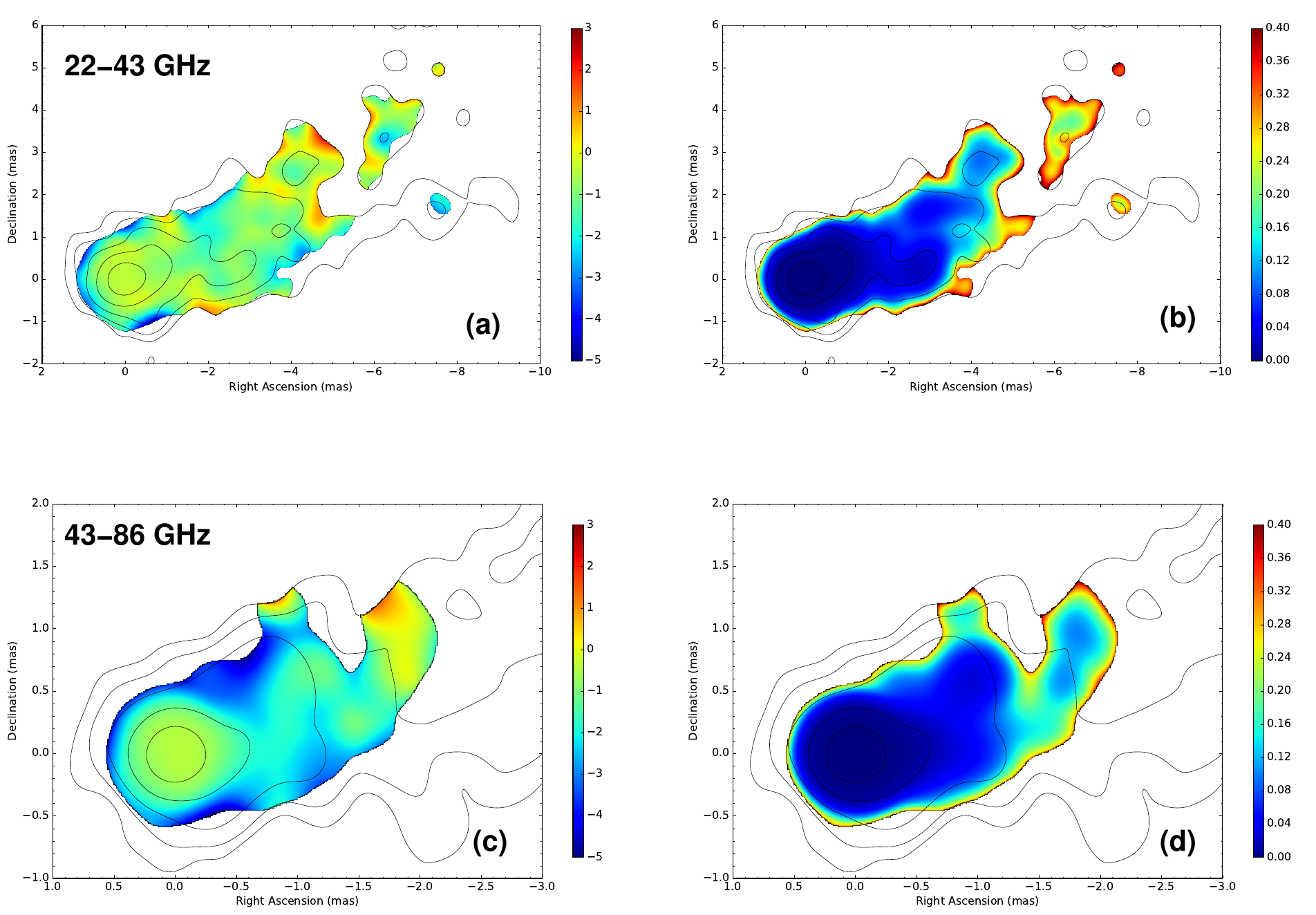}
\caption{The pseudo-color maps of spectral-index distribution
with color bars superimposed on the total intensity 
contour plots. Panel  (\textbf{a}) shows 22--43 GHz spectral-index map superimposed on the contour plot of 22~GHz. 
Panel  (\textbf{c}) shows the 43--86~GHz spectral-index map 
superimposed on the contour plot of 43~GHz. Their noise maps 
showing distribution of uncertainty are also presented in panel  (\textbf{b}) and (\textbf{d}) respectively. }
\label{spe}
\end{figure}

The practical steps are listed as follow:
(1) First, we produce VLBI images using the fully self-calibrated dataset. 
A same $u-v$ range is used to minimize the difference of baseline coverages 
between 22 and 43~GHz. Both images are restored with a circular beam with a 
diameter of 0.65~mas (the same as used in Section~\ref{SSJM}).
(2) Second, we produce an initial spectral-index map by simply aligning the
 geometric center of two images. With~the initial map, we roughly determined 
 the optical thick region which will be masked in the next step.
(3) Third, we import the two images into an interactive Python program, 
VIMAP \citep{KT2014}, based on the algorithm introduced by~\cite{CG2008}. 
A circular mask with a diameter of ~1~mas is used to cover the core region, 
and a rectangle box with relative RA range from 2 to $-$4~mas and DEC range from $-$2 to 
3~mas is used to choose a optically thin region. The~VIMAP gives the values of 
2D-correlation after shifting and find out the shifting corresponding to the 
maximum (see Figure~\ref{2dc}). In~our case, shifting corresponding to the 
maximum is zero. (4) The final spectral-index map is produced in AIPS with 
task COMB. Only regions with intensity $>5\sigma$ are used to make sure the 
result is reliable and a noise map showing the distribution of errors
is also produced. The~22--43~GHz spectral-index map superimposed on the 
intensity contour plots of 22~GHz are presented in panel (a) 
of Figure~\ref{spe} with its noise map in the panel (b).

We also produce a 43--86~GHz spectral-index map with the same procedure.
Considering there is a gap of six days between the observations of two frequencies, 
this spectral-index map is produced only for reference. In~calculating the 
2D cross-correlation, a~circular mask with a diameter of ~0.5~mas and a 
rectangle box with a RA range from 1 to $-$2~mas and a DEC range from $-$1 to 
0.5~mas are used. The~43--86~GHz spectral-index map with its noise map are 
presented in the lower panel of Figure~\ref{spe}. 

On the 22--43~GHz spectral-index map, inverted-spectrum with index $\alpha>0$ ($S_{obs}\propto\nu^{\alpha}$) 
are detected between $d=$~1.30 and 1.60~mas on the edge of the northern limb, 
and between $d=$~1.30 and 1.80~mas on the edge of the southern limb. Some less 
confident inverted-spectrum are seen on the edge of northern limb around
$d$ = 3.80~mas. Strongly inverted spectrum with $\alpha>2$ is detected in 
the northern limb between $d=$~5.00 and 7.00~mas where the limb is interrupted.
These inverted-spectrum features indicate absorptions in the jet in M87.  
Interestingly, on~the map, we see a pattern-like distribution along the jet, 
the regions with steep spectrum interlace with regions with flatter spectrum 
(the pattern-like distribution is also seen on the 43--86~GHz spectral-index map).

\section{Discussion}
\unskip
\subsection{Jet Recollimation with~Absorption}

As we have mentioned in  Section \ref{SSJM}, recollimations between $d=$~0.90 and 1.30~mas 
and between $d=$~3.65 and 4.25~mas could possibly be the representations of the 
persistent recollimations suggested by~\cite{Wa2018} at our observed epoch.
By comparing our obtained transverse jet structure and the 22--43~GHz spectral-index map,
we find these two recollimations are spatially close to the inverted-spectrum 
features. 

The collimations are nonequilibrium behaviors in the AGN jet-launching region 
before the hydrodynamic or magnetohydrodynamic equilibrium is achieved, and~they 
reminds us the reconfinement nodes in the simulations of a jet propagating in the 
external medium with pressure decreasing slower than the thermal pressure of jet 
\citep{Gom1997, Kom1997, Agu2001, Mat2012, Miz2012, Miz2015, Mar2016, Fue2018}. 
As to the external medium, since these features are between 300--2000$R_s$ from 
the central engine (consider the viewing angle of the jet in M87 as 17$^{\circ}$, 
as used in~\cite{Wa2018}), in~the hot accretion flow models like ADIOS (see~\cite{Yu2014} for an overview), the~external medium interacting with an AGN 
jet on this scale is believed to be ``winds'', i.e.,~the moderately magnetized, 
non-relativistic un-collimated, extremely hot and generally fully ionized gas 
outflows launched from the accretion~disk.

Since stationary shocks at the reconfinement node may enhance the magnetic 
fields locally \citep{Miz2015, Mar2016, Fue2018}, thus increase the opacity of 
Synchrotron Self-Absorption, SSA could be a possible mechanism responsible for 
the spectral turnover at the recollimations. But~in a SSA-only case, according 
to Equation~(2) of~\cite{Mar1983}, for~a region with intensity of $0.02~\mathrm{Jy~Beam^{-1}}$ 
(a typical value of intensity of jet in M87 between 300--2000$R_s$), turnover 
frequency higher than 43~GHz requires a unreasonably intense magnetic filed of $10^5~G$. 
So contribution of other mechanisms like free-free absorption from the external medium ("winds") 
must be substantial, therefore spectral turnover at the recollimations may also 
indicate a large free-free opacity, i.e.,~high density of external medium there,
suggesting either the jet collimations locally enhance the density of the external medium,
or the locally high density of external medium induces the collimations of the jet.  
(by looking back upon literatures, we find that the 22--43~GHz spectral-index map in~\cite{Ly2007} 
also showed an association of collimation and absorption at locations between $d=$~1 and 2~mas,
but it was not reported explicitly then). 

\subsection{Jet Interruption with~Absorption}

As we have mentioned in Section \ref{RES}, an~interruption of the northern limb between 5 
and 7~mas from the center of the image is seen at 22~GHz. On~the restored 22~GHz 
image, the~peak of the northern limb around $d=$~6.0~mas is below 5$\sigma$.  
According to the results of 22~GHz KaVA monitoring and 15~GHz MOJAVE survey on M87, 
at many epochs in a period longer than 20 years, an~interruption is seen in the 
northern limb somewhere between $d=$~5 and 10~mas, thus the jet in M87 shows a 
single-ridgeline morphology in this region, e.g.,~at 22~GHz,  2 March 2014 and 
3 May 2014 \citep{Had2017}; at 15~GHz,  1 November 1998, 9 August 2004,  
11 February 2010,  27 August 2019,~et~al. 

Inverted spectrum with $\alpha$ up to 2.2 is detected around the interruption on 
our 22--43~GHz spectral-index map, suggesting a strong absorption feature in this 
area. Interestingly, we find that the inverted spectrum is also seen at a similar 
location around $d=$~7~mas on the 22--43~GHz spectral-index map in~\cite{Ly2007}. 
Since $\alpha$ is very close to the upper limit of SSA (if we set the cutoff to 
3$\sigma$, the~$\alpha$ is almost 3), free-free absorption may probably be the main 
absorption mechanism here. Since this feature is between 2000--4000$R_s$ from the 
central engine, ``winds'' are still believed to be the main absorption medium.
With our intensity slices, we estimate the opacity of this absorption feature: 
on the restored 22 and 43~GHz images, the~intensity peak of the northern limb 
at $d$ = 6.0~mas is $\mathrm{1.5~mJy~beam^{-1}}$ and $\mathrm{2.3~mJy~beam^{-1}}$
respectively. If~emissions were optically thin here, assuming the optically thin 
spectral index is $-$2, then the intensity peak in this region should be 
$\mathrm{8.8~mJy~beam^{-1}}$ at 22~GHz. If~$\mathrm{S_{obs}=S_{pre}e^{-\tau}}$, 
then this implies an opacity of $\mathrm{\tau\sim1.8}$ at 22~GHz. 

The interruption of the northern limb between 2000 and 4000$R_s$ from the 
central engine is an indication that the external medium is not uniform in the 
jet-launching region. As~the jet propagates outward, it may go through regions 
with high f-f opacity. In~addition, the~pattern-like spectral-index distribution 
along the jet might be an indication of non-uniform external medium as well. 
\subsection{Temporal Features and Short-Lived~Instability}

As we have mentioned in Section \ref{SSJM}, on~our 86~GHz image, the~northern limb looks 
rather ragged. The~analysis of transverse structure shows the northern limb starts 
at a larger core distance than its southern counterpart, and~it has a tortuous 
and fragmentized structure, which means its orientation changes several times abruptly, 
resulting in structural features in the jet segment 2, 3, 4 and 6. These phenomenon 
have never been reported by works based on long-term mm-VLBI monitoring, so they might 
just be temporal features, and~represents of the variability of the AGN jet 
in its launching~region.

It has been proved by mm-VLBI observations, that the flow in the jet in M87 is rotating 
clockwisely and with helical magnetic fields \citep{Wa2018}. A~local, internal kink instability 
can grow over short timescales in a jet like this and lead to oscillations \citep{Miz2012, Sin2016, 
Bro2016, Str2016,Bri2017}. At~the observed epoch, a~southward jet oscillation is found 
between $d=$~0.75~mas and 3.00~mas, and~most clearly seen at $d=$~1.45~mas. 
So, the~temporal features could possibly be attributed to the observed southward 
jet oscillation caused by the kink instability. The~northern edge of the jet in M87 was 
compressed by the oscillation, thus local structure of the northern limb was damaged, 
and finally brought this fragmentized morphology of the northern limb and these temporal 
structural features at the observed~epoch.

\section{Conclusions}
With the multi-frequency quasi-simultaneous VLBI observations on M87,
for the fisrt time, we find a spatial association between the jet collimations and the
local enhancement of the density of external medium (``winds'') in the jet-launching region,
suggesting either the jet collimations locally enhance the density of the external medium,
or the locally high density of external medium induces the collimations of the jet.
We also find that the external medium in the jet-launching region,
is not uniform, and~greatly contribute to the absorption in this region.
In addition, we find that for a rotating jet with helical magnetic filed,
like the one in M87, its temporal morphology in the launching region may be
largely affected by the local, short-lived kink instability growing in~itself.

\acknowledgments{We sincerely thank the anonymous 
referee for her/his careful reviewing that improved 
the manuscript. The~VLBA is an instrument of the 
National Radio Astronomy Observatory. 
The National Radio Astronomy Observatory is a facility 
of the National Science Foundation operated under 
cooperative agreement by Associated Universities, Inc.
We also thank Kazuhiro Hada for his valuable suggestions
on the data analysis. This work made use of the Swinburne 
University of Technology software correlator, developed as 
part of the Australian Major National Research Facilities 
Programme and operated under~license.This research is funded by National Natural Science Foundation of China~11803062.}

\authorcontributions{
Conceptualization, X.H., W.Z., T.A., F.W.; data reduction and writing, W.Z.; software and visualization, W.Z., X.L.; doubel-check of data reduction, X.C.; funding acquisition, W.Z.}

\conflictofinterests{The authors declare no conflict of interest.}

%
%

\renewcommand\bibname{References} 

\end{document}